\documentclass[twocolumn]{aastex6}
\usepackage{graphicx}
\usepackage[suffix=]{epstopdf}
\usepackage{natbib}
\usepackage{amsmath}
\usepackage{url}
\usepackage{xspace}

\newcommand{\Kepler}{\textsl{Kepler}\xspace}
\newcommand{\MOST}{\textsl{MOST}\xspace}

\begin{document}

\title{\MOST Observations of our Nearest Neighbor: Flares on Proxima Centauri}

\shorttitle{Flares on Proxima}
\shortauthors{Davenport et al.}

\author{
	James R. A. Davenport\altaffilmark{1,2},
	David M. Kipping\altaffilmark{3}, 
	Dimitar Sasselov\altaffilmark{4},
	Jaymie M. Matthews\altaffilmark{5},
	Chris Cameron\altaffilmark{6}
	}

\altaffiltext{1}{Department of Physics \& Astronomy, Western Washington University, 516 High St., Bellingham, WA 98225, USA}
\altaffiltext{2}{NSF Astronomy and Astrophysics Postdoctoral Fellow}
\altaffiltext{3}{Department of Astronomy, Columbia University, 550 West 120th Street, New York, NY 10027, USA}
\altaffiltext{4}{Harvard-Smithsonian Center for Astrophysics, 60 Garden Street, Cambridge, MA 02138, USA}
\altaffiltext{5}{Department of Physics and Astronomy, University of British Columbia, 6224 Agricultural Road, Vancouver, BC V6T 1Z1, Canada}
\altaffiltext{6}{Department of Mathematics, Physics \& Geology, Cape Breton University, 1250 Grand Lake Road, Sydney, Nova Scotia, Canada, B1P 6L2}

\begin{abstract}
We present a study of white light flares from the active M5.5 dwarf Proxima Centauri using the Canadian microsatellite \MOST.
Using 37.6 days of monitoring data from 2014 and 2015, we have detected 66 individual flare events, the largest number of white light flares observed to date on Proxima Cen.
Flare energies in our sample range from $10^{29}$--$10^{31.5}$ erg,. 
The flare rate is lower than that of other classic flare stars of similar spectral type, such as UV Ceti, which may indicate Proxima Cen had a higher flare rate in its youth. Proxima Cen does have an unusually high flare rate given its slow rotation period, however. 
Extending the observed power-law occurrence distribution down to $10^{28}$ erg, we show that flares with flux amplitudes of 0.5\% occur 63 times per day, while superflares with energies of $10^{33}$ erg occur  $\sim$8 times per year. Small flares may therefore pose a great difficulty in searches for transits from the recently announced 1.27 M$_\earth$ Proxima b, while frequent large flares could have significant impact on the planetary atmosphere. 
\end{abstract}

\section{Introduction}

\begin{figure*}[]
\centering
\includegraphics[width=7in]{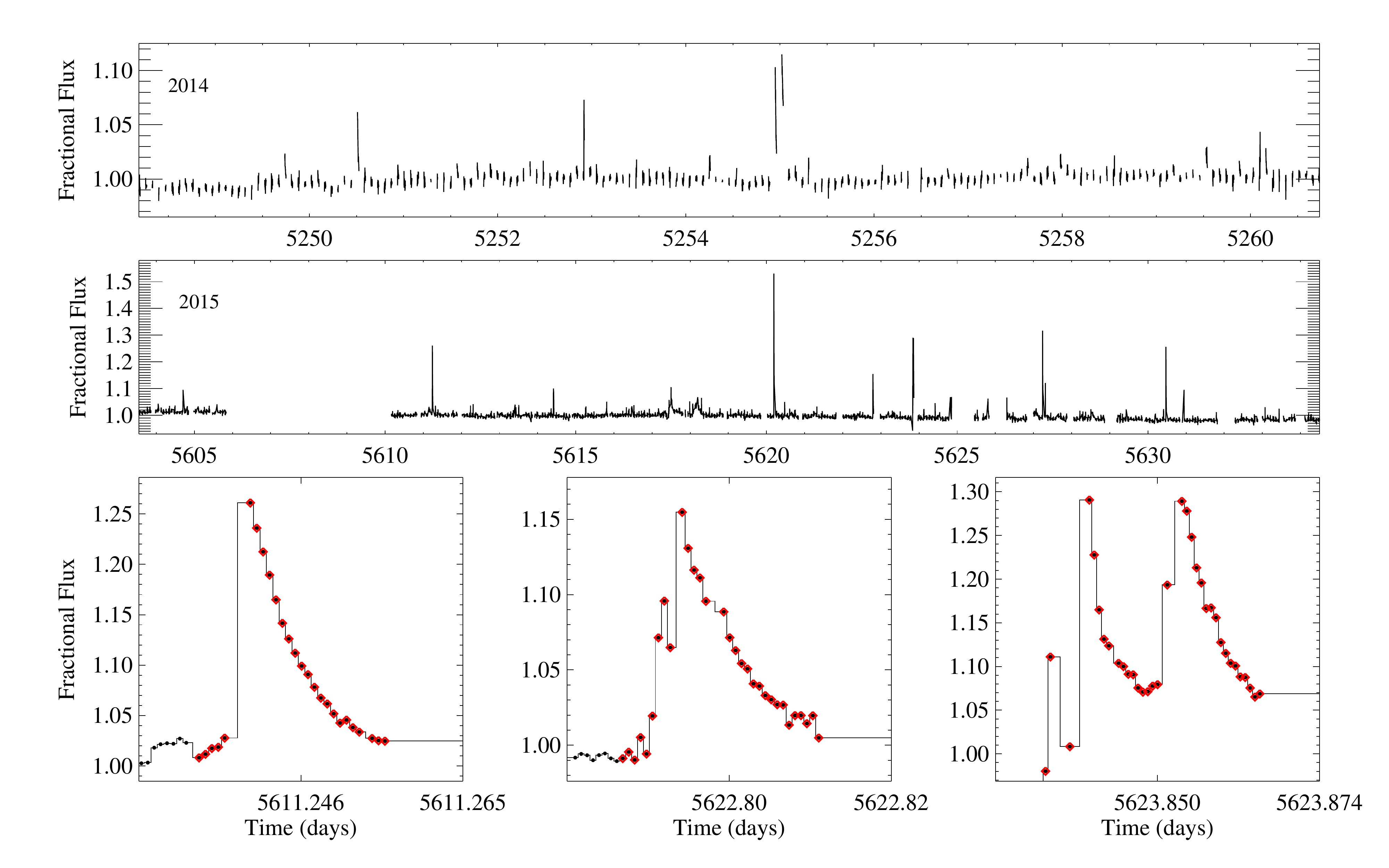}
\caption{Top: Light curve for Proxima Cen obtained by \MOST during the 2014 and 2015 observing runs. Many large amplitude white light flares are visible throughout both observing campaigns. Bottom: Three representative large flares detected on Proxima Cen. Epochs determined to be part of a flare event are highlighted (red points). These flares show a range of morphologies, from ``classical'' to highly complex, multi-peaked structure. In each panel time is relative to HJD(2000) = 2451545 days.}
\label{fig:lc}
\end{figure*}

Stellar magnetic activity presents a challenge to both the habitability and detectability of planets around M dwarfs. Low-mass stars are key targets in modern exoplanet searches due to their high number density in the Galaxy \citep[e.g.][]{henry2006}, and their small radii that yield larger amplitude transit signatures than from Solar-type stars \citep{irwin2009}. Further incentive to study low-mass stars for habitable zone worlds is the recent discovery that half of M dwarfs host a 0.5--1.4 R$_{\earth}$ planet with an orbit shorter than 50 days \citep{dressing2013}. Surface magnetic activity, such as starspots, emission line variability, and flares, remains significant for M dwarfs up to several Gyr old \citep{west2008},
and produces a noise floor for detecting planets in both transit photometry \citep[e.g.][]{oshagh2013} and radial velocities \citep{saar1997,korhonen2015}. Stellar activity has even been the culprit behind false-detections of exoplanets around M dwarfs in high precision data \citep{robertson2014}.

The role of M dwarf magnetic activity on planetary habitability, particularly from X-ray and UV flux due to quiescent emission and flares, is an ongoing area of research \citep{tarter2007,scalo2007,seager2010}. Planetary habitability is impacted by the properties and evolution of both the star and planet \citep[e.g.][]{vidotto2013}.
While models from \citet{segura2010} show that single large M dwarf flares do not pose a great risk to planetary habitability, young planets may be bombarded with much stronger and more frequent flares than previously predicted \citep{armstrong2016}.
 M dwarf activity can strongly affect planetary atmospheres through photoevaporation on Gyr timescales \citep{owen2016,cuntz2016}. Whether this has a net positive or negative impact on atmosphere loss and planetary habitability is debated \citep{luger2015,owen2016}.

Accurate studies of M dwarf magnetic activity in X-rays, UV, and the blue optical are challenging due to the intrinsic faintness of these cool stars at short wavelengths. However, the nearest M dwarf to the Sun, Proxima Cen, has long been known as an active flare star \citep{thackeray1950}, with a moderately strong magnetic field for its spectral type of M5.5 \citep{reiners2008}. At a distance of 1.3 pc, Proxima Cen enables unique characterization of activity in X-ray through optical \citep{gudel2004a,fuhrmeister2011}, and even radio frequencies \citep{lim1996}. 
High time resolution flare data from Proxima Cen has been used to better understand flare heating mechanisms \citep{reale2004,kowalski2016}.
Decades-long spectroscopic monitoring has even revealed a possible activity cycle for Proxima Cen \citep{cincunegui2007}.

Proxima Cen has been the subject of multiple searches for exoplanets \citep[e.g.][]{benedict1999,endl2008}. Recently, \citet{proximab} discovered the existence of Proxima b, a 1.27 M$_\earth$ planet orbiting within the star's habitable zone, using multiple years of high precision radial velocity monitoring. Proxima Cen is therefore an important benchmark object for understanding planet formation around low-mass stars, the evolution of the magnetic dynamo, and the impact of stellar activity on planetary atmospheres.

Using data from the Microvariability and Oscillations of STars microsatellite \citep[hereafter \MOST;][]{walker2003}, we have conducted a study of the white light flares from Proxima Cen. 
Flares have previously been studied with \MOST for the famous active M3.5 dwarf, AD Leo \citep{huntwalker2012}. 
Despite its long history of study as an active M dwarf, the census of flares from Proxima has not been constrained by modern space-based photometric studies. Very few other mid-to-late M dwarfs are bright enough or located within acceptable viewing zones to be studied by similar space-based missions such as \Kepler \citep{borucki2010}. 
These high-precision light curves provide an unparalleled ability to statistically characterize rates and energy distributions for stellar flares \citep[e.g.][]{hawley2014}.

Our data on Proxima Cen come from two observing seasons with \MOST, with 12.7 days of data from 2014 (2648 measurements total), and 24.9 days from 2015 (12762 measurements total). The light curves for both seasons are shown in Figure \ref{fig:lc}. Since Proxima Cen is not in the spacecraft's continuous viewing zone, the 101 minute orbit of \MOST results in periodic gaps in the light curve. A typical cadence of 63.4 seconds was used for both observing seasons, giving a comparable temporal resolution to \Kepler short-cadence observations. Data reduction was carried out using the typical procedure for \MOST data \citep{rowe2006}.

\section{Identifying White Light Flares}

The reduced light curves were analyzed in fractional flux units, as shown in Figure \ref{fig:lc}, and had typical uncertainties of $\sim$0.5\%. 
A $\sim$3\% decay in brightness was found over the 31 day span of the 2015 observing season, which may be due to the slow rotational signature of Proxima \citep{benedict1998b,benedict1999}, and is consistent with the 8\% flux modulation seen in \citet{proximab}.
However, no de-trending or pre-whitening was performed to remove this slow stellar variability, as it did not affect our ability to detect flares or measure their energies.

The \MOST light curves from both the 2014 and 2015 observing seasons were analyzed with v1.3.12 of the flare-finding suite {\tt FBEYE}\footnote{\url{https://github.com/jradavenport/FBEYE}} from \citet{davenport2014b}, which was originally developed for flares studies with \Kepler data. 
This {\tt IDL} software package provides a simple automatic flare-finding algorithm that identifies data points above a running smoothed light curve, as well as a graphical user interface for inspecting light curves and vetting the detected flares. 
Flare start- and stop-times were first identified using the auto-finding prescription, and then adjusted by eye to ensure spurious points or data gaps were not included in flare events. 

Each flare's energy was calculated in {\tt FBEYE} as the Equivalent Duration \citep[e.g. see][]{huntwalker2012}, computed as the trapezoidal sum of the flare in relative flux units above the local quiescent level ($\Delta F / \bar{F}$). 
Due to the regular gaps in the light curve from the spacecraft's orbit, large amplitude, long duration flares were not entirely monitored through to their return to quiescence. This can be seen in the complex flares of Figure \ref{fig:lc}. Our flare energies therefore are lower-limits in these cases, as we did not attempt to reconstruct the missing flare light curves, and we do not include these events in our rate analysis.

\section{Determining Flare Energies}

\begin{figure}[]
\centering
\includegraphics[width=3.25in]{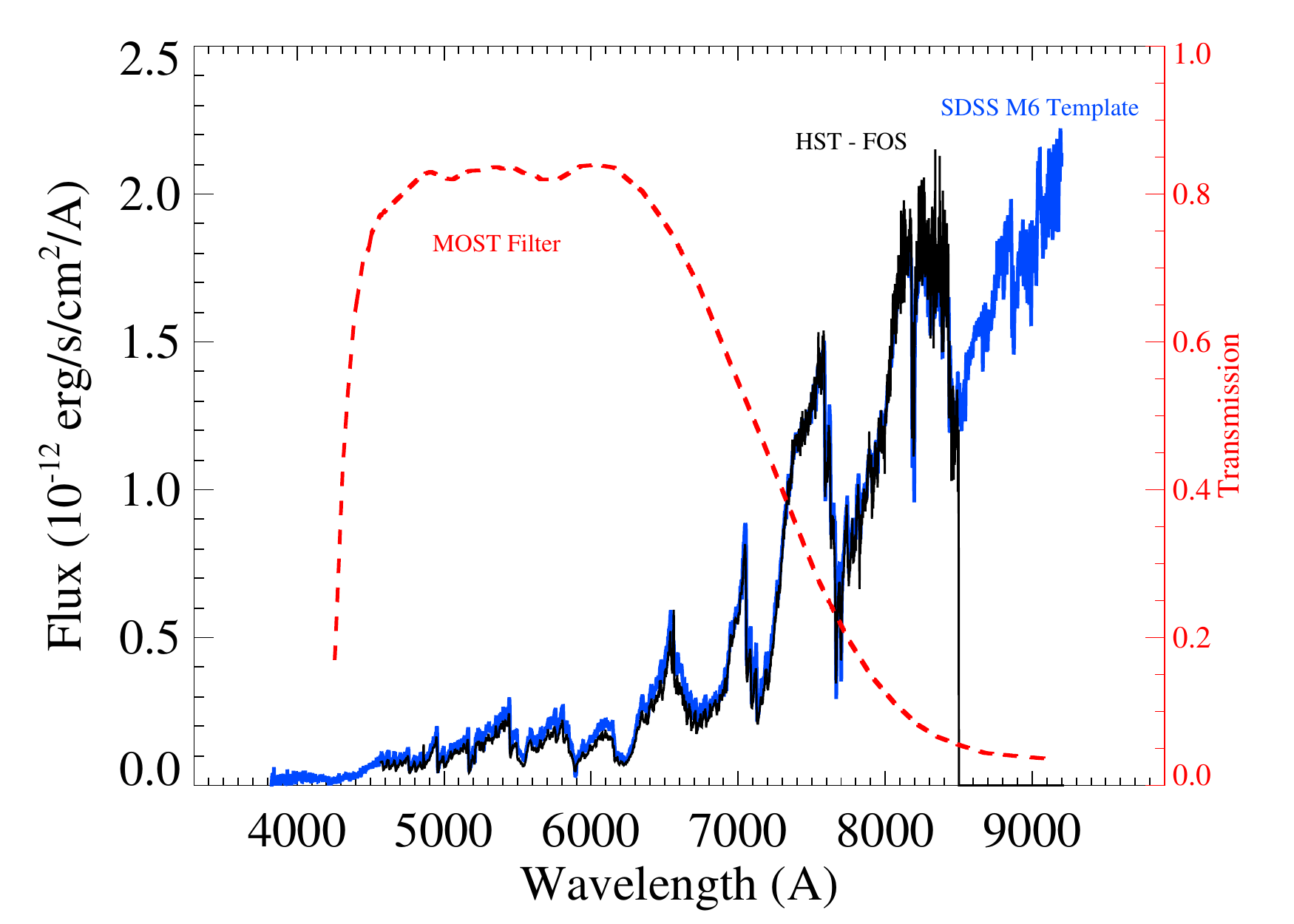}
\caption{
Proxima Cen spectrum from the HST Faint Object Spectrograph (black line), along with the scaled \citet{bochanski_templates} M6 spectral template (blue line). Overlaid is the \MOST filter transmission curve (red dashed line).
}
\label{fig:spec}
\end{figure}

In order to convert the equivalent durations reported by {\tt FBEYE} to physical energies for each flare event, we needed to determine the quiescent luminosity of Proxima Cen through the \MOST bandpass. We assumed a distance for Proxima Cen of $d=1.3018 \pm 0.0002$ pc (parallax of 768.13 mas per year) from \citet{lurie2014}. The response function for the \MOST filter from Fig. 4 of \citet{walker2003} is also reproduced in Figure \ref{fig:spec}. 

A  flux calibrated quiescent spectrum for Proxima Cen was obtained by \citet{schultz1998} using the HST Faint Object Spectrograph. However, this spectrum did not span the full range of the \MOST filter. To fill in these gaps, we scaled the M6 spectral template from  \citet{bochanski_templates} to the flux-calibrated HST spectrum. The composite spectrum is shown in Figure \ref{fig:spec}. Convolving this final spectrum with the \MOST filter curve, we determined a quiescent luminosity for Proxima Cen in this bandpass of log L$_0$ = 28.69 erg s$^{-1}$. Finally, to compute flare energies we multiplied the equivalent duration (in units of seconds) by L$_0$.

\section{Flare Statistics}

\begin{figure}[]
\centering
\includegraphics[width=3.5in]{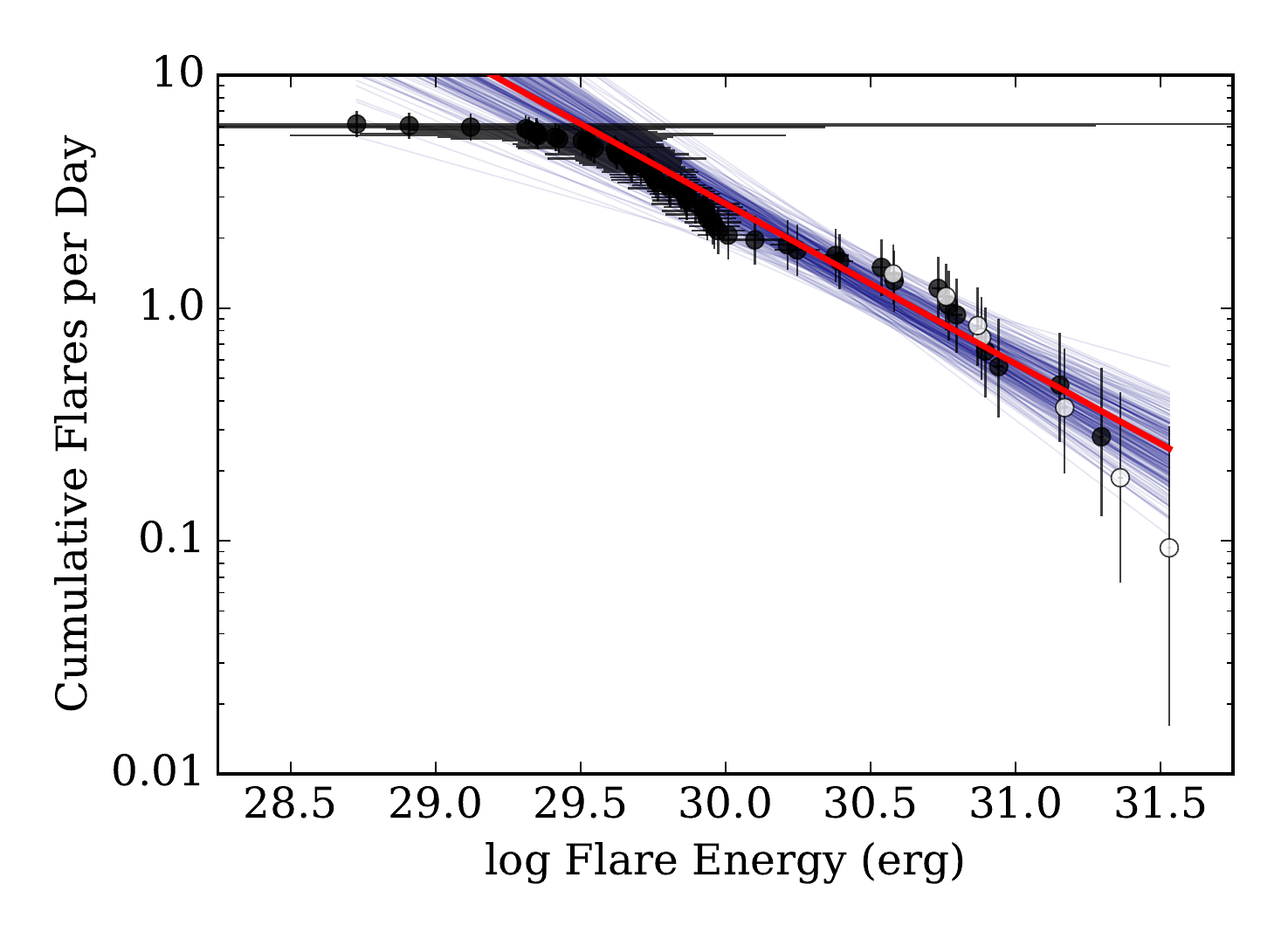}
\caption{
Cumulative flare frequency distribution for the 50 events observed on Proxima Cen with \MOST (filled black points), with uncertainties on both the frequency and flare energy (grey bars). 
A power-law was fit to the data (red line) using the Bayesian MCMC approach from \citet{kelly2007}. 250 draws from the posterior distribution of this fit are shown (blue lines).
Long duration flares whose measured energies are only lower limits were not used in the fit (open circles).
}
\label{fig:ffd}
\end{figure}

A total of 66 flare events were identified between the two observing seasons, with 15 from 2014 and 51 from 2015. The total flare rate (determined as the number of flares in the observing season divided by the total exposure time) was constant in both seasons to within the Poisson uncertainty, with  $8.1\pm2.7$ and $5.7\pm0.9$ flares per day in 2014 and 2015, respectively. In total, 7.5\% of the observed data (1159 epochs) was classified as flaring.

In Figure \ref{fig:ffd} we present the cumulative flare frequency distribution (FFD) versus event energy, which is the typical parameter space used to characterize stellar flare rates. Flare frequency uncertainties were calculated using the Poisson confidence intervals \citep{gehrels1986}, while errors for the flare energies were computed as the inverse signal to noise ratio \citep[e.g. see][]{lurie2015}. The apparent saturation of the flare rate at low energies is due to incompleteness in recovering low amplitude events. 
A power-law was fit to this distribution with weighting in both the energy and frequency dimensions, using the Markov Chain Monte Carlo (MCMC) approach of \citet{kelly2007}. We used 10,000 steps in our MCMC chains, finding a best-fit power-law of 
$\log \nu = - 0.68 (\pm 0.10) \log E  + 20.9( \pm 3.2)$.

The previous best estimate of optical flare statistics for Proxima Cen came from 35 flares events observed with U-band over 4 days by \citet{walker1981}. Our sample contains flares more than an order of magnitude higher in energy than found by \citet{walker1981}. They recovered a FFD slope of -0.69, very close to our fit in Figure \ref{fig:ffd}. However, \citet{walker1981} found a lower cumulative rate of flares, with $\sim$0.5 flares per day at a representative flare energy of $\log E_U$ = 30.5 erg, compared to our power-law fit that yielded 1.2 flares per day at the same energy in the \MOST band. 
While the U-band flare energies are not directly comparable to those from \MOST, \citet{hawley2014} showed a flare in the similar \Kepler-band had $\sim$1.5$\times$ more flux than the U-band for the active M4 dwarf GJ 1243. Assuming this same transformation was roughly applicable to Proxima Cen and the \MOST filter, our measured flare rate at the equivalent $\log E_U$ = 30.5 erg would be $R_{30.5}\sim$0.9 flares per day, somewhat closer to the previous constraint.

Interestingly the FFD for Proxima Cen shown in Figure \ref{fig:ffd} is very close to that of GJ 1245 B from \cite{lurie2015}, who found both components of the M5+M5  GJ1245 AB binary system had remarkably similar flare rates using 9 months of \Kepler data. Both components of GJ 1245 AB have rotation periods shorter than 1 day, implying a young age for the system. 
The most robust rotation period for Proxima Cen is 83.5 days from \citet{benedict1998b}, with spot modulations up to $\sim$4\% found using HST FGS data, though \citet{kippingmost} also find shorter periodicities in the \MOST photometry with an unclear origin. 
The similar flare rates between Proxima Cen and GJ 1245 B, despite two orders of magnitude difference in rotation period, is a challenge to rotation--activity relationships for flares, such as those found with \Kepler for G through early M dwarfs \citep{davenport2016}.

Other famous M5--M6 stars with well studied flare rates include CN Leo, and the prototypical flare star, UV Ceti \citep{lme1976}, both of which show higher overall rates of flares in the U-band compared to the results from \citet{walker1981}. Our improved flare census places Proxima Cen slightly higher than CN Leo in total flare rate, but a factor of 2 lower than UV Ceti. Table \ref{tbl} provides flare rates and rotation periods for several other stars with similar spectral type to Proxima Cen.
For stars without robust rotation periods from starspot monitoring we have computed the rotation period as $2\pi R_\star / v \sin i$.
Additional rotation periods for active stars with spectral types of M5 or later are desperately needed to further explore the evolution of the magnetic dynamo and flare rates for fully convective flare stars.

\begin{deluxetable}{lccc}
\tablecolumns{4}
\tablewidth{0pt}
\tabletypesize{\footnotesize}
\tablecaption{Compilation of rotation periods and cumulative flare rates evaluated at a fixed energy of $10^{30.5}$ erg for several well-studied flare stars with a similar spectral type to Proxima Cen.
\label{tbl}}
\tablehead{
	\colhead{Star}&
	\colhead{Spectral Type}&
	\colhead{$P_{rot}$ (days)}&
	\colhead{$R_{30.5}$ (\#/day)}
	}
\startdata 
GJ 1245 A & M5 & 0.26d\tablenotemark{a} & 2.3\tablenotemark{a} \\
GJ 1245 B & M5 & 0.71d\tablenotemark{a}& 1.1\tablenotemark{a} \\
Prox Cen & M5.5 & 83.5d\tablenotemark{b} & 1.2\tablenotemark{$\star$} \\
EQ Peg\tablenotemark{$\dagger$} & M5 & 0.52\tablenotemark{c} &38\tablenotemark{d}\\
CN Leo & M5 & $>3$\tablenotemark{c} & 0.6\tablenotemark{d} \\
UV Ceti\tablenotemark{$\dagger$}& M5.5 & 0.26\tablenotemark{c} & 3.1\tablenotemark{d} 
\enddata
\tablenotetext{\tablenotemark{$\star$}}{This work}
\tablenotetext{\tablenotemark{$\dagger$}}{Binary system}
\tablenotetext{a}{\citet{lurie2015}}
\tablenotetext{b}{\citet{benedict1998b}}
\tablenotetext{c}{Using $v \sin i$ from \citet{mohanty2003}}
\tablenotetext{d}{\citet{lme1976}}
\end{deluxetable}

\begin{figure}[]
\centering
\includegraphics[width=3.5in]{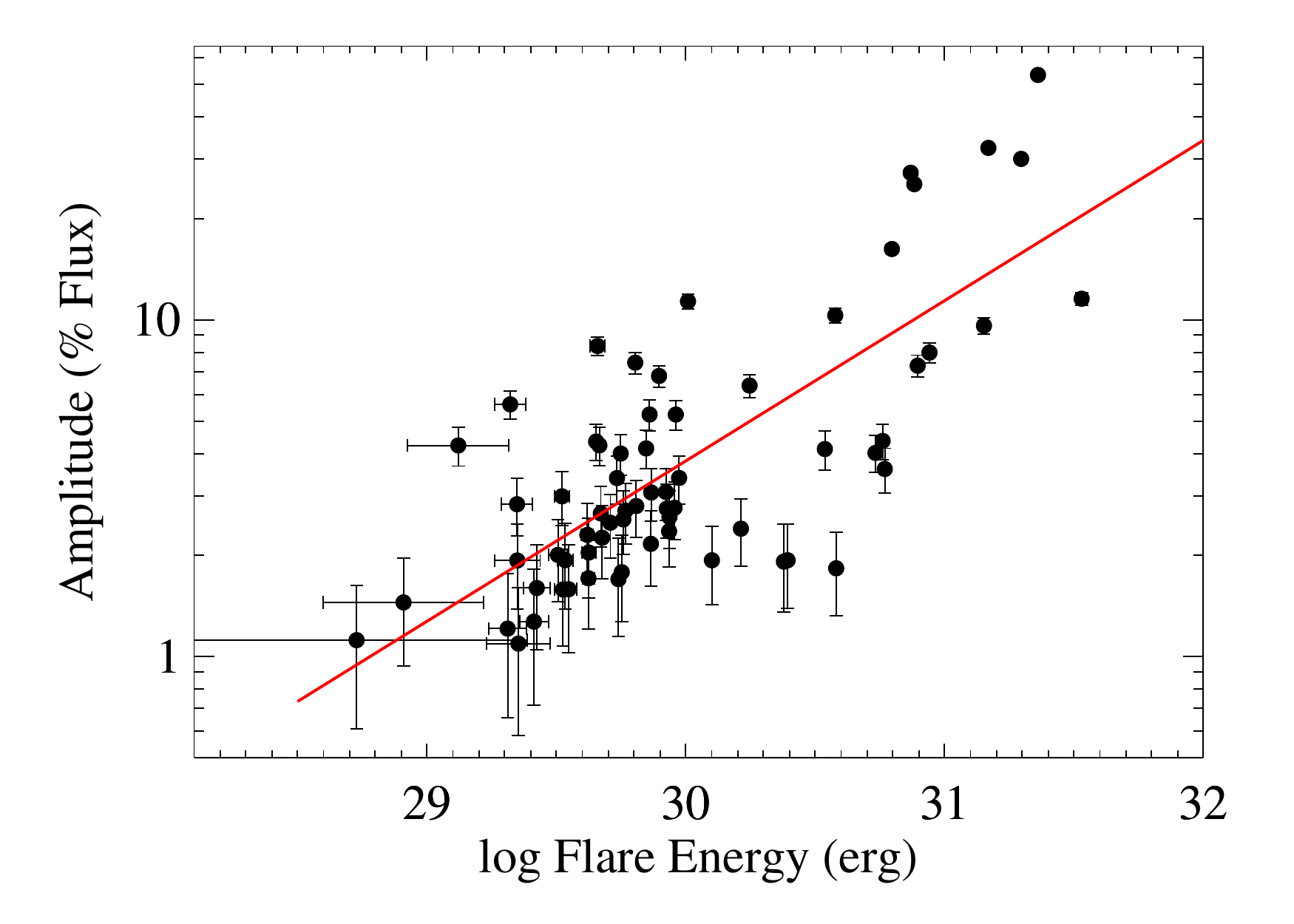}
\caption{Peak amplitudes versus event energies for the 50 flares in our sample (black points), with accompanying power-law fit (red line). Error bars for the flare amplitudes were computed as the mean photometric uncertainty within each event.}
\label{fig:ampl}
\end{figure}

In Figure \ref{fig:ampl} we show the measured amplitudes versus energies for our entire sample of flares.  Five flares had amplitudes greater than 0.25 mag. The flare peak amplitude transformations from \citet{davenport2012} indicate these large events would have $u$-band fluxes of 2--3 mag.
A power-law was fit, again using the Bayesian MCMC approach from \citep{kelly2007} with uncertainties on both the amplitudes and event energies, and had the form 
$\log A = 0.48 (\log E) - 13.6$. 
This relation indicates that a flare with log E $\sim$ 28 would have an amplitude of 0.45\%, comparable to that of a transiting 1 R$_\earth$ exoplanet in the habitable zone of Proxima Cen \citep[e.g.][]{nutzman2008}. Extrapolating the cumulative rates from Figure \ref{fig:ffd} down to this energy, we predict that Proxima Cen produces 63 flares per day with an amplitude of $\sim$0.5\%.

\section{Implications for Habitability}

\citet{proximab} recently announced the discovery of Proxima b, a 1.2 M$_\earth$ planet orbiting Proxima Cen at 0.0485 AU. To fully understand the impact flares have on the habitability of a planet with such a close-in orbit, detailed models of photochemistry and photoevaporation are required. 
The seminal work by \citet{segura2010}, for example, modeled the response by a habitable zone terrestrial planet's atmosphere to a large M dwarf flare. This simulation found life-threatening UV flux only reached the planetary surface for $<$100 sec, and the atmosphere recovered fully within 2 years.

The large flare in \citet{segura2010} had a total white light energy of log E $\sim$ 34 erg, and was based on the real event from AD Leo (M3.5) observed by \citet{slhadleo}. The largest flares in our sample have log E $\sim$ 31.5 erg, more than 2 orders of magnitude smaller in energy. However, \citet{lurie2015} found using much longer time-series from \Kepler that the FFD for M5 stars can reach energies of at least log E $\sim$ 33 erg. Extending our FFD fit from Figure \ref{fig:ffd}, we predict Proxima Cen could produce $\sim$8 flares at log E = 33 erg per year.

Proxima b is located squarely in the habitable zone for Proxima Cen \citep{endl2008}, 3.3 times closer than the 0.16 AU assumed for the Earth-analog around AD Leo by \citet{segura2010}. This results in a $\sim$10 times higher insolent flare flux for a given event compared with AD Leo. Given a maximum flare energy for an M5 star of log E $\sim$ 33 erg, 1 dex lower than the event from AD Leo, our flare rate suggests Proxima b could experience a comparable flare event to that studied in \citet{segura2010} multiple times per year.  If these flares regularly impacted Proxima b, the atmosphere would never fully recover. 
While this is not known to be a ``show-stopper'' for habitability, it clearly necessitates a more detailed investigation of atmospheric response on minutes to years timescales, and photoevaporation over Gyr timescales for Proxima b.

\section{Summary and Discussion}

We have presented a census of the white light flare activity for Proxima Cen, finding 66 events from 37.6 days of monitoring over two seasons. Flares in our sample span more than 2 orders of magnitude in energy. While Proxima Cen does not exhibit an exceptional number of flares for an active M dwarf, it does show unusually high flare activity given its slow rotation period. 
\citet{wright2016a} have recently shown fully convective stars appear to follow the same rotation--activity evolution as Solar-type stars. This discrepancy in flare rates indicates that some external agent is driving heightened flare activity on Proxima Cen, rather than a breakdown in the rotation--flare activity evolution for fully convective stars. 
Star-planet tidal interactions due to Proxima b are an unlikely cause for the increased stellar activity, given gravitational perturbations from Proxima b on Proxima Cen are 3--4 orders of magnitude lower than is expected to affect the stellar dynamo \citep{cuntz2000}.

With the recently announced discovery of Proxima b, the flare activity for this star is especially interesting. Our flare rate indicates Proxima Cen could produce $\sim$8 superflares per year at its present age, and 63 flares per day with amplitudes comparable to the transit depth expected for Proxima b. Comparing our flare rate to other M5--M6 stars suggests Proxima was more active in its youth. The current flare rate is quite similar to that of GJ 1245B, despite Proxima Cen having a very slow rotation period.

Though our data cannot constrain the properties of coronal mass ejections (CMEs) from Proxima Cen associated with flares, these eruptions may have a large impact on the habitability of Proxima b \citep{khodachenko2007}. As large CME opening angles are associated with high energy flare events \citep{taktakishvili2011}, Proxima b would be frequently bombarded by these coronal ejections, which may greatly impact the survival and composition of any planetary atmosphere \citep[e.g.][]{barnes2016proxima}. Proxima Cen is thus a high priority target for spectroscopic monitoring to study CME velocities and rates associated with flare events \citep[see e.g.][]{vida2016}.

\acknowledgments
The authors thank the anonymous referee for their helpful comments that improved this manuscript. 
JRAD is supported by an NSF Astronomy and Astrophysics Postdoctoral Fellowship under award AST-1501418.

Based on data from the \MOST satellite, a Canadian Space Agency mission, jointly operated by Microsatellite Systems Canada Inc. (MSCI; formerly Dynacon Inc.), the University of Toronto Institute for Aerospace Studies and the University of British Columbia, with the assistance of the University of Vienna.

\end{document}